\documentclass{ws-procs9x6}

\begin{document}

\title{
MULTIPOLE ANALYSIS OF RECENT PION ELECTROPRODUCTION DATA
}

\author{
R. Arndt, W. Briscoe, I. Strakovsky, and R. Workman
}

\address{
Center for Nuclear Studies, Department of Physics \\
The George Washington University, Washington, D.C. 20052 \\
}

\maketitle

\abstracts{
We have performed a series of global and single-Q$^2$ fits 
to the growing database for pion electroproduction.  We 
compare to some other recent fits which have used different 
formalisms, more selective datasets, and truncated multipole 
expansions.  We present results for the E2/M1 and S2/M1 
ratios, as a function of Q$^2$. 
}

\section{Fit Procedure}

In fitting the pion electroproduction database~\cite{menu}, 
we have fixed the Q$^2$ = 0 point based on our fits to pion 
photoproduction~\cite{sm02}.  The photoproduction multipoles 
have been parametrized using the form
\begin{equation}
M\; = \; ({\rm Born} + \alpha_B ) ( 1 + i T_{\pi N} ) +
\alpha_R T_{\pi N} + ( {\rm Im} T_{\pi N} - T_{\pi N}^2 ) 
(\alpha_r^H + i \alpha_i^H ) ,
\label{eq:photo}
\end{equation}
containing the Born terms, with phenomenological pieces 
($\alpha$) maintaining the correct threshold behavior 
and Watson's theorem.  The $\pi$N T-matrix (T$_{\pi N}$) 
connects each multipole to structure found in the elastic 
$\pi$N scattering analysis.  Additional structure is 
parametrized through the Born terms (which contain
vector-meson contributions), and the phenomenological 
structure functions ($\alpha_B$, $\alpha_R$, $\alpha_r^H$, 
and $\alpha_i^H$).

At non-zero Q$^2$, the Born terms have built-in Q$^2$ 
dependence, other terms have been modified by a factor
\begin{equation}
f(Q^2) = { k \over {k(Q^2=0)}} {1\over {(1+Q^2 / 0.71)^2}}
e^{- \Lambda Q^2} \left( 1 + \alpha Q^2 + \beta 
\left[ {W\over {W_R}} -1 \right] \right) ,
\label{eq:ff}
\end{equation}
where k is the photon CM momentum, $\Lambda$ is a 
universal cutoff factor, and the constants, $\alpha$ and 
$\beta$, are searched for each multipole.  We have 
performed energy and $Q^2$-dependent fits over the full 
kinematic range [W to 1.7~GeV and Q$^2$ to 5~(GeV/c)$^2$ ] 
in addition to fitting data clustered around particular 
Q$^2$ values.  

\begin{figure}[th]
\centerline{\psfig{file=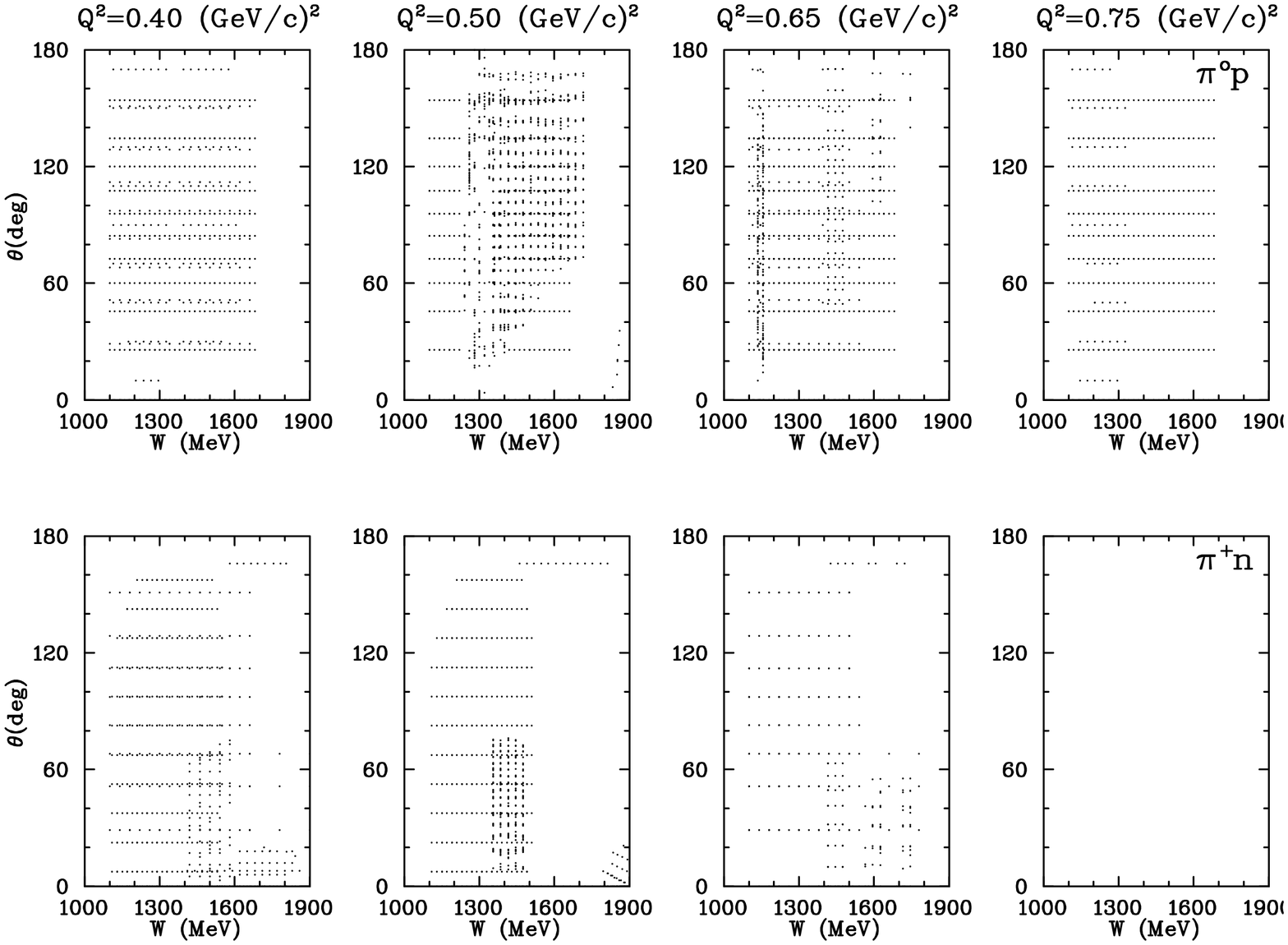,width=10cm,clip=,silent=,angle=90}}
\vspace{0.2in}
\centerline{\psfig{file=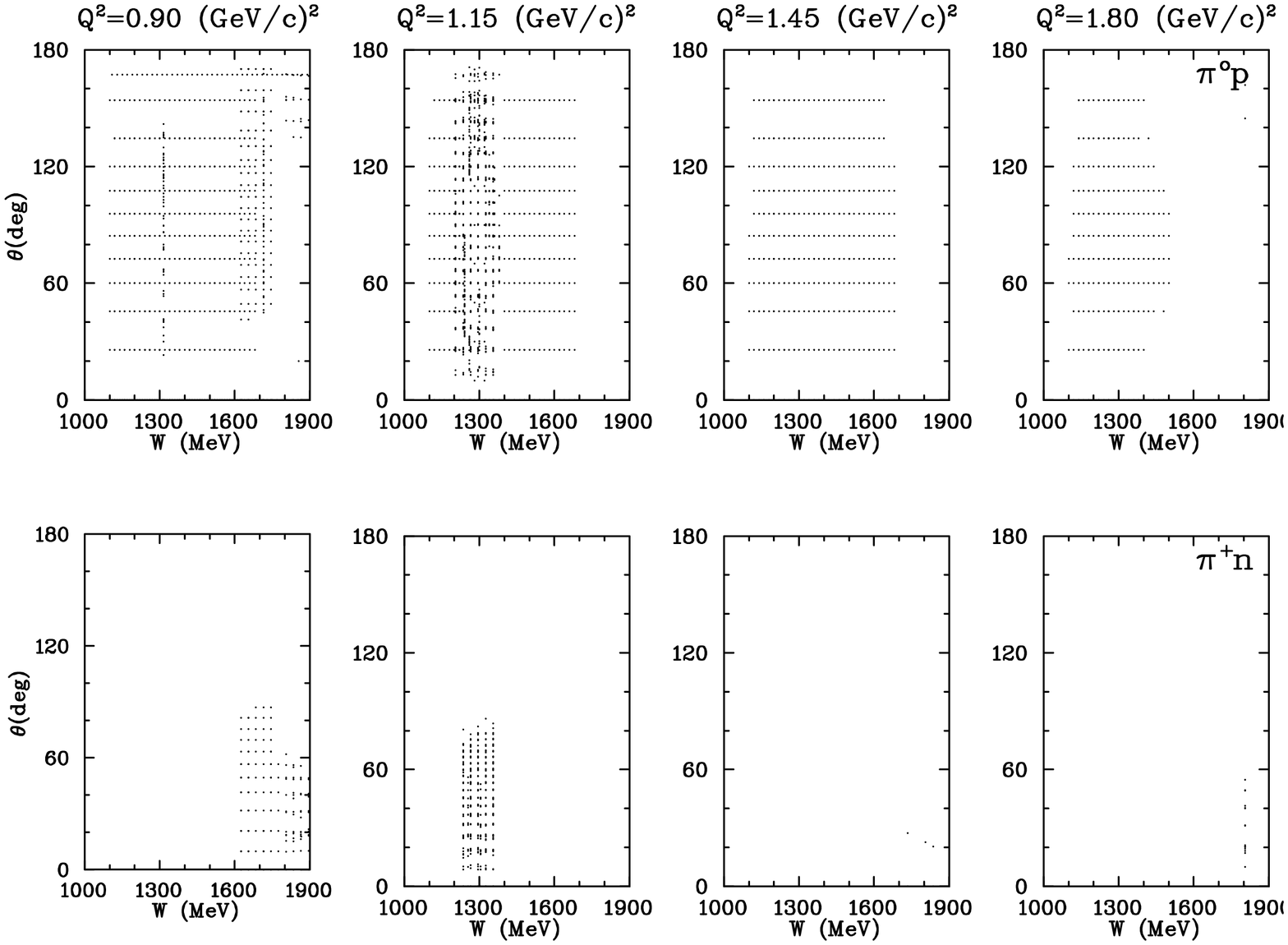,width=10cm,clip=,silent=,angle=90}}
\vspace*{8pt}
\caption{Energy-angle distribution of $\pi^0$p and $\pi^+$n 
         electroproduction data: Q$^2$ = 0.40, 0.50, 0.65, 
         0.75, 0.90, 1.15, 1.45, and 1.80~(GeV/c)$^2$ 
         associated with JLab CLAS $\pi^0$p Q$^2$ binning 
         for single-Q$^2$ results. \label{fig1}}
\end{figure} 
\begin{figure}[th]
\centerline{\psfig{file=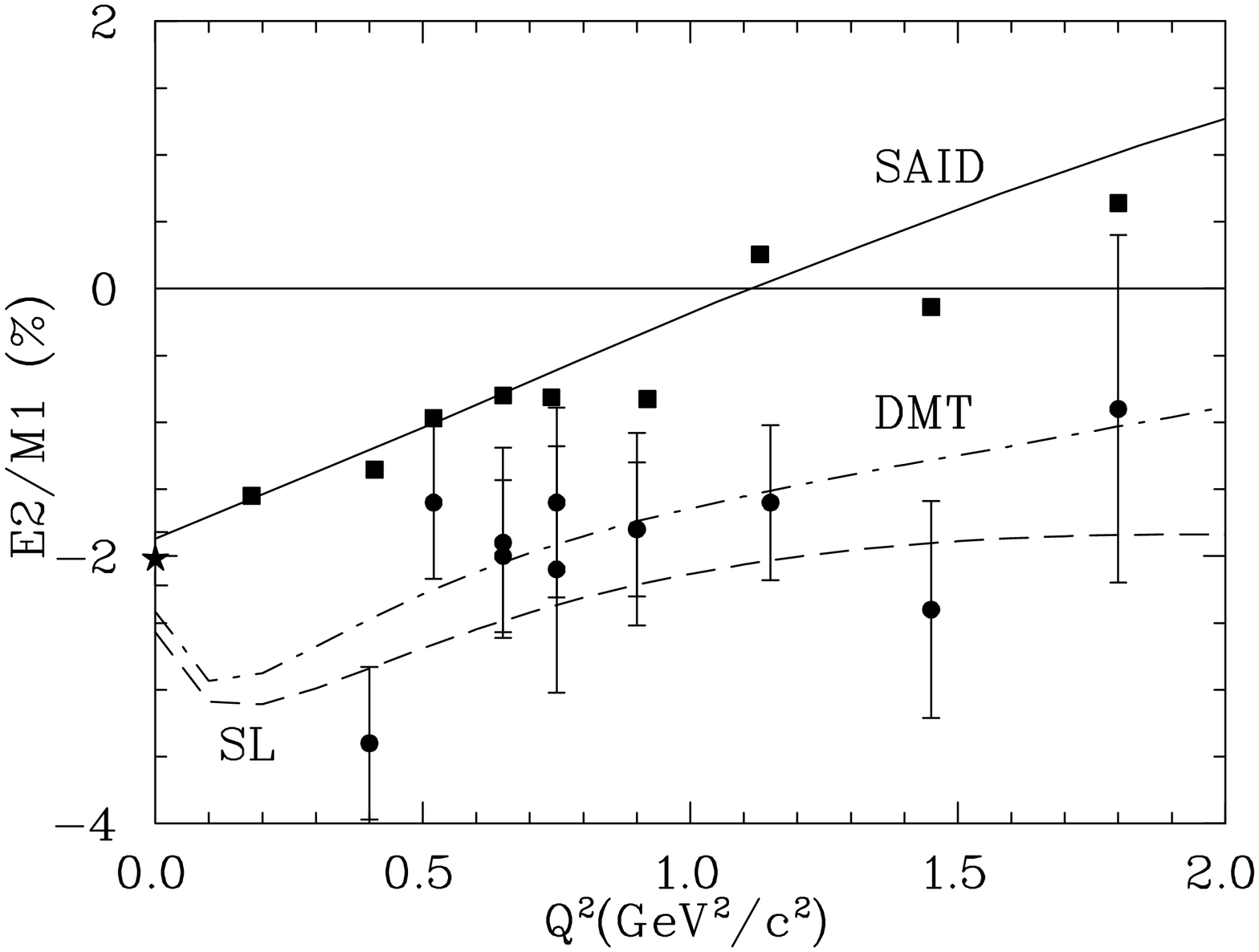,width=5.5cm,clip=,silent=,angle=90}\hfill
            \psfig{file=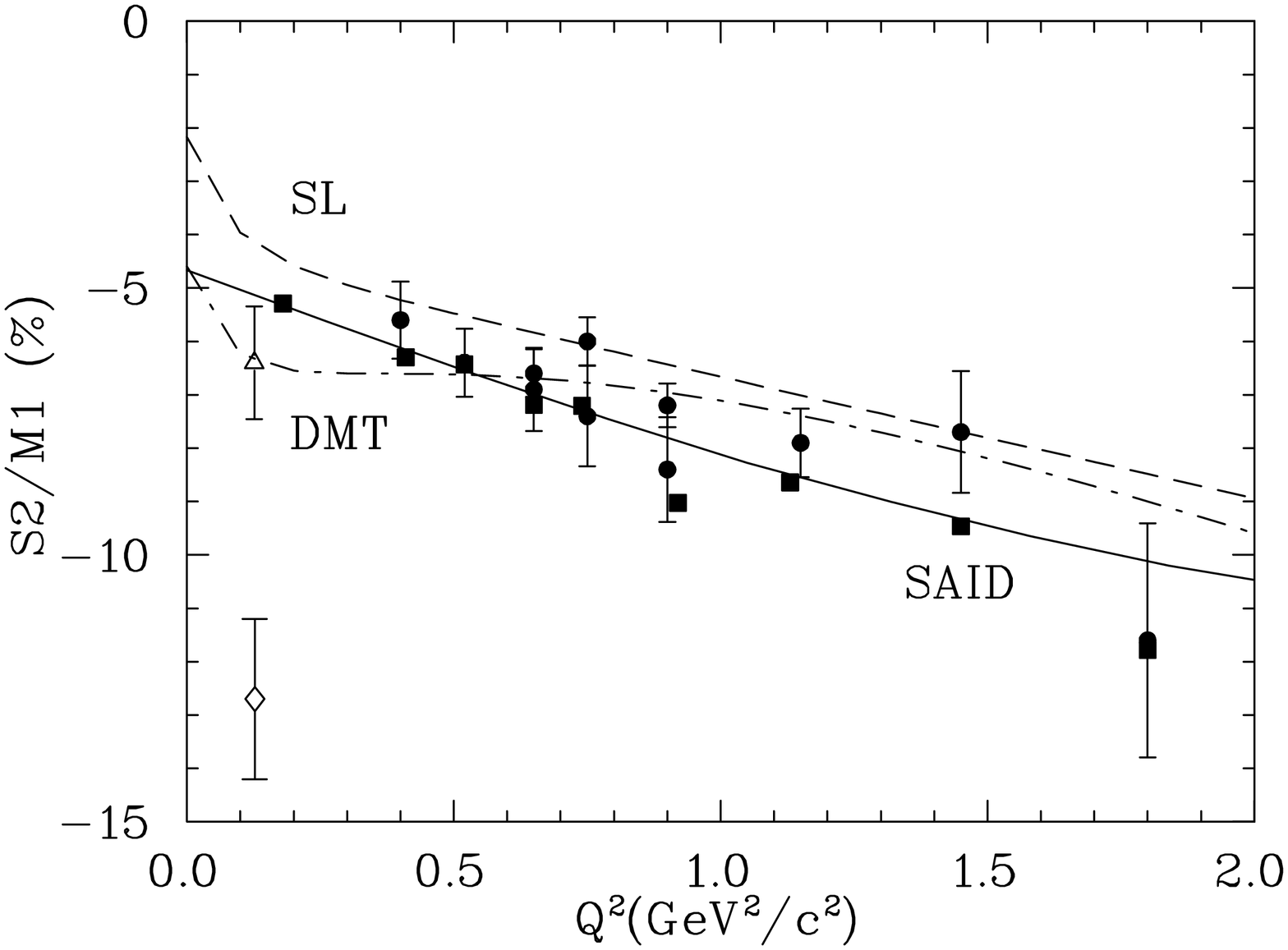,width=5.5cm,clip=,silent=,angle=90}}
\vspace*{8pt}
\caption{E2/M1 and S2/M1 ratios \textit{vs} Q$^2$.  
         Values were extracted from our fixed Q$^2$ 
         analyses starting on the global fit (filled 
         squares) using world and JLab CLAS data~
         \protect\cite{clas}.  Results from  Ref.~
         \protect\cite{clas} (filled circles) are 
         given.  In addition, (left panel), our pion 
         photoproduction result (Q$^2$ = 0)~
         \protect\cite{sm02} (filled asterisk), and 
         (right panel), the data of Refs.~
         \protect\cite{Mainz01} (open triangle) and
         \protect\cite{Bonn97} (open diamond) are 
         shown.  The solid curves give our global 
         (energy and Q$^2$ dependent) best-fit 
         results.  Dash-dotted and dashed curves are 
         from Refs.~\protect\cite{maid} and 
         \protect\cite{Lee}, respectively.\label{fig2}}
\end{figure}
\begin{figure}[t!]
\leftline{\psfig{file=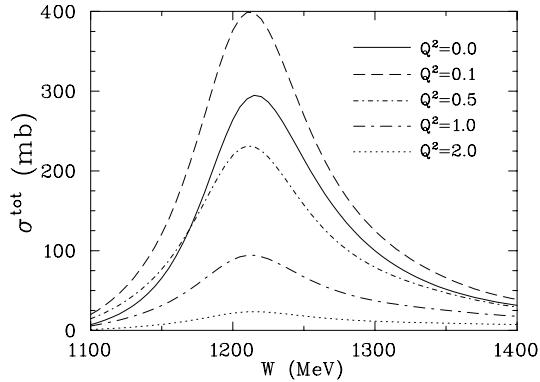,width=7cm,clip=,silent=,angle=90}}
\vspace{-5mm}
\hspace*{.65\textwidth}\raisebox{30mm}[0pt][0pt]
{\parbox{.35\textwidth}{\caption[fig1]{\label{fig3}Total
         cross section for $\pi^0$p electroproduction, showing
         the variation with $Q^2$.}}}
\end{figure}
\begin{figure}[t!]
\leftline{\psfig{file=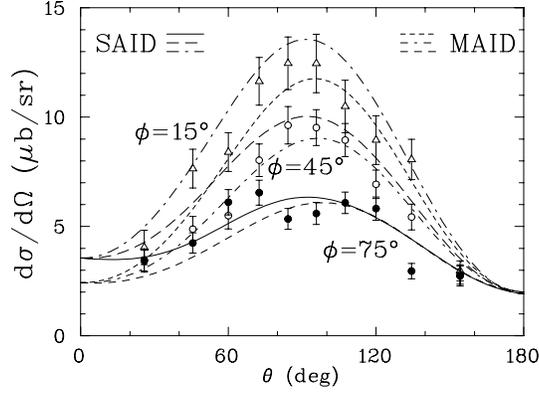,width=7cm,clip=,silent=,angle=90}}
\vspace{-5mm} 
\hspace*{.65\textwidth}\raisebox{30mm}[0pt][0pt]
{\parbox{.35\textwidth}{\caption[fig1]{\label{fig4}
         Differential cross section for $\pi^0$p 
         electroproduction, showing the $\phi$ dependence.
         Data are from ~\protect\cite{clas}.}}}   
\end{figure}
\begin{figure}[th]
\centerline{\psfig{file=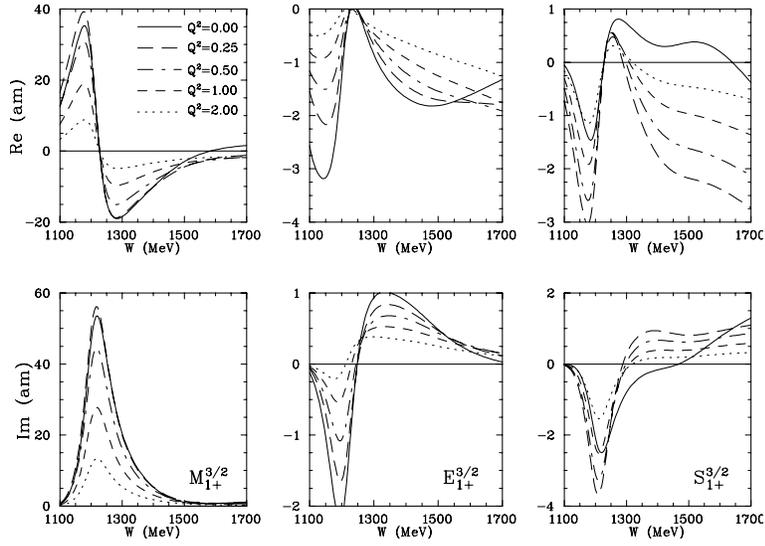,width=10cm,clip=,silent=,angle=90}}
\vspace*{8pt} 
\caption{Q$^2$ dependence of the magnetic (M$_{1+}$), 
         electric (E$_{1+}$), and scalar (S$_{1+}$) 
         multipoles, for isospin 3/2. \label{fig5}}
\end{figure}

As can be seen in Figure~\ref{fig1}, there is relatively 
little $\pi^+n$ data at moderate $Q^2$.  As a result, 
some analyses actually quote a ratio of E2/M1 for the 
$\pi^0p$ charge channel, rather than the isospin 3/2 
component, assuming a negligible isospin 1/2 part at 
the $\Delta$-resonance.

\subsection{Data Reduction}

A fit to distributions in $\theta$ and $\phi$ can be made 
more efficient if the symmetry $\sigma (\phi )$ = $\sigma 
(360 - \phi )$ is used.  Let $\sigma_1$ = $\sigma (\phi )$, 
$\sigma_2 = \sigma ( 360 - \phi )$, and z be the 
statistical error (squared).  If we take $\sigma$ to be 
the fitted value, then for each pair of $\phi$ values we 
have
\begin{equation}
\chi^2 ={{ \left(  \sigma - \sigma_1 \right)^2 } \over z_1 } 
     +  {{ \left(  \sigma - \sigma_2 \right)^2 } \over z_2 },
\label{eq:ch1}
\end{equation}
which can be written as
\begin{equation}
\chi^2 = {{ \left( \sigma_1 - \sigma_2 \right) }^2 \over {z_1 + z_2} }
    +  {{ \left( \sigma - \sigma_a \right) }^2 \over z_a },
\label{eq:ch2}
\end{equation}
where $\sigma_a = (z_1\sigma_2 + z_2\sigma_1)/(z_1 + z_2)$ 
is the weighted average of $\sigma_1$ and $\sigma_2$, 
and z$_a$ is given by
\begin{equation}
{1 \over {z_a}} = { 1 \over {z_1}} + { 1 \over {z_2}}.
\label{eq:ch3}
\end{equation}  
Using the form given in Eq.~(\ref{eq:ch2}), only half 
of the data points are required in the fitting routine, 
the first term being a constant which measures the 
internal consistency of a given set.  For the set of 
$\pi^0$p data given in Ref.~\cite{clas}, for example, 
the first term in Eq.~(\ref{eq:ch2}) contributes a 
$\chi^2$ of 15040 for the 12528 $\phi$ pairs.   

\section{Results}

In Figures~~\ref{fig2}$-$~\ref{fig5}, we compare our result 
to MAID~\cite{maid}, Sato-Lee~\cite{Lee}, and single-$Q^2$
fits performed by the Jefferson Lab CLAS~\cite{clas} 
Collaboration.  Below 2~(GeV/c)$^2$, there is considerable 
scatter, with SAID and MAID extrapolating to an E2/M1 
zero-crossing; no crossing being seen in Sato-Lee.  As 
indicated above, the higher-Q$^2$ region is sparsely 
populated.  In our fits, we find that the inclusion of 
additional preliminary data can shift the crossing point 
significantly, but a crossing always occurs.  This is in 
particular disagreement with the 4~(GeV/c)$^2$ Hall~C 
value~\cite{Frolov}, and details of the various fits, 
at this kinematic point, are being studied in more 
detail. 

\section*{Acknowledgments}

This work was supported in part by a U.~S.~Department of 
Energy Grant No.~DE--FG02--99ER41110.  We also acknowledge 
a contract from Jefferson Lab under which part of this 
work was done.  Jefferson Lab is operated by the 
Southeastern Universities Research Association under DOE 
contract DE--AC05--84ER40150.

\end{document}